# Non linear Prediction of Antitubercular Activity Of Oxazolines and Oxazoles derivatives Making Use of Compact TS-Fuzzy models Through Clustering with orthogonal least sqaure technique and Fuzzy identification system

Doreswamy

Department of Computer Science

Mangalore University,

Mangalagangotri-574 199, Karnataka, INDIA

Doreswamyh@yahoo.com

Chanabasayya .M. Vastrad

Department of Computer Science

Mangalore University,

Mangalagangotri-574 199, Karnataka, INDIA

channu.vastrad@gmail.com

*Abstract*-The prediction of uncertain and predictive nonlinear systems is an important and challenging problem. Fuzzy logic models are often a good choice to describe such systems, however in many cases these become complex soon. commonlly, too less effort is put into descriptor selection and in the creation of suitable local rules. Moreover, in common no model reduction is applied, while this may analyze the model by removing redundant data. This paper suggests a combined method that deal with these issues in order to create compact Takagi-Sugeno (TS) models that can be effectively used to represent complex  predictive systems. A new fuzzy clustering method is come up with for the identification of compact TS-fuzzy models. The best relevant consequent variables of the TS model are choosen by an orthogonal least squares technique based on the obtained clusters. For the selection of the relevant antecedent (scheduling) variables a new method has been developed based on Fisher's interclass separability basis. This complete approach is demonstrated by means of the  Oxazolines and Oxazoles derivatives as antituberculosis agent for nonlinear regression benchmark. The results are compared with results obtained by neuro-fuzzy i.e. ANFIS algorithm and advanced fuzzyy clustering techniques i.e FMID toolbox .

## I. INTRODUCTION

Fuzzy modeling and identification has become an important research area because fuzzy models are capable of handling perceptual uncertainties, such as the unclearness and ambiguity involved in a real system, and has shown excellent ability when describing nonlinear system on the basis of observed data[1,2].  The most frequently applied Takagi-Sugeno (TS) model tries to  decompose the input space of the nonlinear model into fuzzy  subcluster and then approximate the system in each subcluster by a simple linear regression model [3,4]. Moreover, it is suitable to be applied in model-based prediction, for the consequent function is the affine linear model. Due to its outstanding features, T–S fuzzy model has been utilized in new drug discovery applications.

Different ways to obtain TS-fuzzy models from data have been presented. Most methods, however, utilize only the function approximation techniques of fuzzy systems, and little attention is paid to the qualitative aspects. This makes them less suited for applications in which emphasis is not only on accuracy, but also on interpretability, computational complexity and maintainability[5]. Additionally, such direct approaches soon lead to quite analyzable models because generally little effort is put into descriptor selection and in the creation of suitable local rules. Moreover, in general no model reduction is applied, while this may simplify the model by removing redundant descriptors. This paper presents a combined method that handles these issues in order to create compact Takagi-Sugeno (TS) models that can be effectively used to represent complex systems.

The bottleneck of the TS model identification is the data-driven identification of the structure that requires nonlinear optimization. For this purpose often cut and try, data-driven approaches, like fuzzy clustering approaches are applied, mainly for determining the rule preexistent of TS models[6-7]. We come up with a more advanced clustering approach that moves a further step in accomplishing the total parameter and structure identification of TS models. This approach is based on a new clustering algorithm called the modified Gath-Geva clustering[8]. The clusters are created by expectation maximization (EM) identification of the TS model in the form of a mixture of Gaussians model [9].

The acquired model is then reduced by reducing the quantity of antecedent variables and also the quantity of consequent variables. Utilizing as well many antecedent variables may result in problems in the prediction and interpretability capabilities of the model due to redundancy, non-informative descriptors and outliers. For that reason, selection of the scheduling variables is frequently necessary. For this intention, we modify our procedure that is based on Fischer interclass separability procedure and have been developed for descriptors selection of fuzzy clusters[10-11].

An Orthogonal Least Squares (OLS) procedure is designed for reduction of the consequent space. The purpose of orthogonal transforms for the reduction of the number of rules has taken much attention in current literature [12-13]. These procedures evaluate the output contribution of the rules to obtain an significance ordering. For modeling intentions, the OLS is the most appropriate tool. In this paper, OLS is applied for a different intention; the selection of the most relevant input and consequent variables based on the OLS analysis of the local models of the clusters.

The objectives of this study were to develop a Takagi–Sugeno fuzzy cluster model with OLS method to predict antitubercular activity from Oxazolines and Oxazoles derivatives descriptor dataset using cross validation approach , compare model performance with a Fuzzy Model Identification (FMID) toolbox that is based on Gustafson-Kessel clustering [14-15] and MATLAB fuzzy toolbox (ANFIS, neuro-fuzzy model)[16-17].

## II. MATERIALS AND ALGORITHAMS

### A. The Data Set

The molecular descriptors of 100 Oxazolines and Oxazoles derivatives [18-19] based H37Rv inhibitors analyzed. These molecular descriptors are generated using Padel-Descriptor tool [20]. The dataset covers a diverse set of molecular descriptors with a wide range of inhibitory activities against H37Rv. The pIC50(observed biologcal activity) values range from -1 to 3. The dataset can be arranged in data matrix. This data matrix u contains m samples(molecule structures) in rows and k descriptors in columns. Vector y with order m×1 denotes the measured activity of interest i.e pIC50. When modeling, both u and y are mean-centered.

### B. Fuzzy Model Identification Toolbox

Fuzzy model identification is an effective tool for the approximation of uncertain nonlinear systems on the basis of measured data. The identification of a fuzzy model using input-output data can be divided into two tasks: structure identification, which determines the type and number of the rules and membership functions, and parameter identification. For both structural and parametric adjustment, prior knowledge plays an important role. Hence, in this book the rules of the fuzzy system are designed based on the available a priori knowledge and the parameters of the membership, and the consequent functions are adapted in a learning process based on the available input-output data. Hence, this chapter is devoted mainly to the parameteridentification of the proposed fuzzy models, but certain structure identification tools are also discussed.

### C. Adaptive neuro fuzzy inference system (ANFIS)

ANFIS is a kind of neural network that is based on Takagi–Sugeno fuzzy inference system. Since it integrates both neural networks and fuzzy logic principles, it has potential to capture the benefits of both in a single framework. Its inference system corresponds to a set of fuzzy IF–THEN rules that have learning capability to approximate nonlinear functions [21]. Hence, ANFIS is considered to be universal approximator [22].

### D. Mathematical approach for the applied Takagi-Sugeno models

Our objective is to develop an algorithm for the identification of unknown nonlinear systems of the form:

$$y_k = f(U_k) \quad \text{----------------------} \quad (1)$$

based on generated descriptor data $U_k = [u_{1,k} \ldots \ldots u_{m,k}]^T$ and measured activity data $y_k$ of the system, where $K = 1 \ldots N$ denotes the index of the K-th descriptor-activity data pair.

In general it may not be easy to find a global nonlinear model that is totally applicable to describe the unknown system $f(.)$. In that case it would surely be beneficial to build local linear models for perticular operating points of the process and combine these into a global model. This can be finished by combining a number of local models, where each local model has a predefined operating region where the local model is accurate. This results in the so called operating regime based model[23] or more perticuar in a TS-fuzzy model when the operating region is defined by fuzzy rules. This type of operating regime based model is defined as:

$$\hat{y}_k = \sum_{i=1}^{c} \omega_i(X_k)(a_i^T \phi_k + b_i) \text{------------------} (2)$$

where $\omega_i(X_k)$ defines the operating regime of the i-th local linear model defined by the parameter vector $\theta_i = [a_i^T b_i]^T$, $\phi_k$ is the regression vector, where $X_k$ and $\phi_k$ are n and $n_r$ dimensional subsets of the original descriptor vector $U_k$, acordingly. The operating regime of the local models can also be depicted by fuzzy sets[24]. Hence, the whole global model can be easily dipicited by Takagi-Sugeno fuzzy rules[4]:

$R_i$ : If $x_k$ is $A_i(x_k)$ then $\hat{y}_k = a_i^T \phi_k + b_i$, $[w_i]$ i=1,……,c. -------------------(3)

$A_i(x)$ corresponds to a multivariate membership function that describes the fuzzy set $A_i$, where $a_i$ and $b_i$ are the parameters of the local linear model, and $w_i = [0,1]$ is the weight o f the rule that corresponds to the desired impact of the rule (note that w is not ω in eq (2)). The value of $w_i$ is frequently chosen by the designer of the fuzzy system based on his or her certinity in the quility and accuracy of the i-th rule. When such information is not available $w_i$ is set as $w_i = 1, \forall i$. Generaly, the antecedent proposition "$x_k$ is $A_i(x_k)$" is framed as a logical unification of simple propositions with univariate fuzzy sets descrbed for the distinctive components of x, often in the conjunction form:

$R_i$ : If $x_{1,k}$ is $A_{i,1}(x_{1,k})$ and . . . . and $x_{n,k}$ is $A_{i,n}(x_{n,k})$ then $\hat{y}_k = a_i^T \phi_k + b_i$, $[w_i]$ ----- (4)

In this case, β the degree of completin of a rule is calculated as the product of the degree of completion of the fuzzy sets in the rule:

$$\beta_i(x_k) = w_i A_i(x_k) = w_i \prod_{i=1}^{n} A_{i,j}(x_{j,k}) \text{---------} (5)$$

The rules of the fuzzy model are combined using the normalized fuzzy mean formula:

$$\hat{y}_k = \frac{\sum_{i=1}^{c} w_i \beta_i(x_k)(a_i^T \phi_k + b_i)}{\sum_{i=1}^{c} w_i \beta_i(x_k)} \text{----------------------} (6)$$

Gaussian membership functions are used here to express the fuzzy set $A_{i,j}(x_{j,k})$:

$$A_{i,j}(x_{j,k}) = \exp\left(-\frac{1}{2}\frac{(x_{j,k} - v_{i,j})^2}{\sigma_{i,j}^2}\right) \text{--------------} (7)$$

where $v_{i,j}$ express the center and $\sigma_{i,j}^2$ the variance of the Gaussian function. The use of Gaussian membership function admits for the compact formation of eqn (5):

$$\beta_i(x_k) = w_i A_i(x_k) = w_i \exp\left(-\frac{1}{2}(X_k - V_j^x)^T (F_i)^{-1}(X_k - V_j^x)\right) \text{------} (8)$$

where $V_j^x = [v_{1,j}, \ldots, v_{n,j}]$ designates the center of the i-th multivariate Gaussian and $F_i$ poses for a diagonal matrix that includes the variances. In the coming sections a new clustering-based method for the identification of the above granted model structure is explained. In addition we explain new methods for the selection of the $\phi_k$ consequent and the $X_k$ antecedent variables.

### E. Clustering for the identification of TS model

The goal of clustering is to division the identification data Z into c clusters. This way, each observation consists of the decriptor and the activity data, grouped into a row vector $Z_k = [u_k\ y_k]$ where k stands the k th row of the Z matrix. The fuzzy partition is denoted by the $U = [\mu_{i,k}]_{c \times N}$ matrix, where the $\mu_{i,k}$ essential feature of the matrix denotes the degree of membership, how the $Z_k$ observation is in the cluster i = 1,....,c.

The present new fuzzy clustering technique is make use of that has been trained from the Gath-Geva technique[8]. Each cluster includes an input distribution, a local model and an output distribution to represent the density of the data:

$$p(y,u) = \sum_{i=1}^{c} p(y, u, \eta_i) = \sum_{i=1}^{c} p(y, u|\eta_i) = \sum_{i=1}^{c} p(y|\eta_i) p(x|\eta_i) p(\eta_i) \text{ ------ (9)}$$

The input distribution is parameterized as an unconditioned Gaussian[25], and describes the field of influence of a cluster similarly to multivariate membership functions. The clustering is based on the minimization of the sum of weighted squared distances between the data points, $Z_k$ and the cluster prototypes, $\eta_i$ that includes the parameters of the clusters. The square of the $D_{i,k}^2$ distances are weighted with the membership values $\mu_{i,k}$ in the objective function that is underrate by the clustering algorithm and planed as:

$$J(Z, U, \eta) = \sum_{i=1}^{c} \sum_{j=1}^{N} (\mu_{i,k})\ D_{i,k}^2 \text{ --------------- (10)}$$

To attain a fuzzy partitioning space, the membership values have to gladden the following conditions:

$U \in \mathbb{R}^{c \times N} | \mu_{i,k} \in [0\ 1], \forall i, k; \sum_{i=1}^{c} \mu_{i,k} = 1, \forall k; 0 < \sum_{k=1}^{c} \mu_{i,k} < N, \forall i \text{ ------- (11)}$

The minimization of the eq.10 functional act as a non-linear optimization problem that is subject to constrains denoted by eq.11 and can be deal with by using a variety of available techniques. The most popular method, is the one after the other optimization (OO), which is planed as follows:

### E.A Clustering algorithm

*Initlization* Given a set of data Z cite c, select weighting exponent m = 2 and a cut-off limit $\varepsilon > 0$. Initialize the $U = [\mu_{i,k}]_{c \times N}$ partition matrix randomly, where $\mu_{i,k} = p(\eta_i|\phi_k)$ means the membership that the $Z_k$ decriptor-activity data created by the i th cluster.

*Loop* for l = 1,2,......

*Step 1)* Compute the parameters of the clusters

• Compute the centers and standard deviation of the Gaussian membership functions:

$$V_i^{x(l)} = \frac{\sum_{k=1}^{N} \mu_{i,k}^{l-1} x^k}{\sum_{k=1}^{N} \mu_{i,k}^{l-1}}, \quad \sigma_{i,j}^{2(l)} = \frac{\sum_{k=1}^{N} \mu_{i,k}^{l-1}(x_{j-k} - v_{j-k})^2}{\sum_{k=1}^{N} \mu_{i,k}^{l-1}}, 1 \leq i \leq c \text{ -------------- (12)}$$

• Evaluate the parameters of the local models by weighted least squares,

The weighted, also called local, parameter evaluation approach does not evaluate all parameters accordingly, the parameters of the local models are evaluated separately, using a set of local evaluation basis eqn.(4).

$$\min_{\theta_i} = \|y - \Phi_e \theta_i\|^2 \beta_i = \|y\sqrt{\beta_i} - \Phi_e \sqrt{\beta_i} \theta_i\|^2 (13)$$

Where $\Phi_e$ means the augumented regression matrix acquired by adding a unitary column to $\Phi$, $\Phi_e = [\Phi\ 1]$, where $\Phi = [\phi_1, \ldots, \phi_K]^T$ and $y = [y_1, \ldots, y_N]^T$, and $\beta_i$ means a diagonal matrix having membership degrees in its diagonal elements.

$$\beta_i = \begin{bmatrix} \mu_{i,1} & 0 & \cdots & 0 \\ 0 & \mu_{i,2} & \cdots & 0 \\ \vdots & \vdots & \ddots & \vdots \\ 0 & 0 & 0 & \mu_{i,N} \end{bmatrix} \text{-----------------} (14)$$

The weighted least-squares measurement of the consequent rule parameters is given by:

$$\theta_i = (\Phi_e^T \beta_i \Phi_e)^{-1} \Phi_e^T \beta_i y \text{------------------------} (15)$$

• $\sigma_i$ standard deviation of the modeling error:

$$\sigma_i^2 = \frac{\sum_{k=1}^N (y_k - f_i(\phi_k,\theta_i))^T (y_k - f_i(\phi_k,\theta_i)) \mu_{i,k}^{l-1}}{\sum_{k=1}^N \mu_{i,k}^{l-1}} \text{------}(16)$$

• A priori probability of the cluster and the weight (impact) of the rules:

$$p(\eta_i) = \frac{1}{N} \sum_{k=1}^N \mu_{i,k}, \quad w_i = p(\eta_i) \prod_{j=1}^N \frac{1}{\sqrt{2\pi\sigma_{i,j}^2}} \text{-}(17)$$

*Step 2)* Calculate the distance measure $D_{i,j}^2(x_k, \eta_i)$

The distance measure is consists of two terms. The first term is based on the geometrical distance between the $V_i$ cluster centers and $X_k$, while the second is based on the achievement of the local linear models:

$$\frac{1}{D_{i,j}^2(x_k,\eta_i)} =$$

$$w_i \prod_{j=1}^n \exp\left(-\frac{1}{2}\frac{(x_{j,k}-v_{i,j})^2}{\sigma_{i,j}^2}\right) \frac{\exp\left(-(y-[\phi_k\ 1]\theta_i^T)^T(y-[\phi_k\ 1]\theta_i^T)\right)}{2\pi\sigma_{i,}^2}$$
---------------- (18)

*Step 3)* Revise the partition matrix:

$$\mu_{i,k}^{(l)} = \frac{1}{\sum_{j=1}^c (D_{i,k}(Z_k\ \eta_i)/D_{j,k}(Z_k\ \eta_j))^{2/(m-1)}} \quad 1 \leq i \leq c,$$

$$1 \leq k \leq N \text{--------------------------} (19)$$

**until** $\|U^{(l)} - U^{(l-1)}\| < \varepsilon$

Chooseing the antecedent and consequent variables

Utilizing excessive antecedent and consequent variables results in complication in the prediction and interpretability efficiency of the fuzzy model due to redundancy, non-informative features and outliers. To prevent these problems in this section two techniques are shown.

*F. Choosing the Consequent Variables by Orthogonal Least Squares technique*

As the fuzzy model is linear in the parameters $\theta_i$ Eqn.(13) is determined by least squares technique ( see Eqn.(15)) that can be also framed as:

$$\theta_i = B^+ \sqrt{\beta_i} \text{------------------------} (20)$$

Where $B^+$ represents the Moore-Penrose pseudo inverse of $\Phi_e \sqrt{\beta_i}$. The OLS technique converts the coloumns of **B** into a set of orthogonal basis vectors in order to examine the unique contribution of each rule. To do this Gram-Schmidt orthogonalization of **B** = **WA** is utilized where W is an orhogonal matrix $W^T W = I$ and **A** is an upper triangular matrix with unity diagonal elements. If $W_i$ represents the i-th coloumn of W and $g_i$ correlative element of the OLS solution vector $g = A\theta_i$, the output variance $(y\sqrt{\beta_i})^T (y\sqrt{\beta_i})/N$ can be expressed by the regressors $\sum_{i=1}^{n_r} g_i W_i^T W_i/N$. Thus, the error reduction ratio, $\varrho$, because an indidual rule i can be represened as

$$\varrho^i = \frac{g_i^2 W_i^T W_i}{(y\sqrt{\beta_i})^T (y\sqrt{\beta_i})} \sqrt{\beta_i} \text{--------------------} (21)$$

This ratio expreses a simple mean for ordering the consequent variables, and can be easily used to choose a subset of the descriptor's in a forward regression manner.

*F.A Chooseing the Scheduling Variables based on Interclass Separability*

Descriptor selection is mainly essential. For this goal, we modify the Fischer interclass separability technique which is based on statistical properties of the data and has been applied for descriptor selection of labeled data [26]. The interclass separability basis is based on the $F_B$ between-class and the $F_W$ within-class covariance matrices that sum up to the total covariance of the training data $F_T$, where:

$$F_W = \sum_{i=1}^{c} p(\eta_i) F_i, \quad F_B = \sum_{i=1}^{c} p(\eta_i)(V_i - V_0)^T (V_i - V_0), \quad V_0 = \sum_{i=1}^{c} p(\eta_i) \text{------- (22)}$$

The fisher interclass seperatibility selection basis is a compromise between $F_W$ and $F_B$:

$$J = \frac{\det F_B}{\det F_W} \text{--------------------- (23)}$$

The significance of a descriptor is measured by leaving out the descriptor and computing J the reduced covariance matrices. The descriptor selection is made iteratively by leaving out the least needed descriptor.

### G. Measures of Model Performance

Proposed model performance is determined in the same manner for the fuzzy rule-based model and the regression model. Each model was evaluated using two different measures of root mean square error (RMSE) and two different $r^2$ (square of correlation coefficient) measures. Each of the measures of model performance corresponds to each collection of data. A low RMSE and a high $r^2$ are indicative of a model that performs well.

Training-RMSE: This calculates the average square root of the sum of squared errors associated with the training data set.

Test-RMSE: This calculates the average square root of the sum of squared errors during the LOO cross-validation procedure. This sum of squares from the LOO procedure is also referred to as a predicted residual sum of squares statistic.

Training-$r^2$: This is a correlation of predicted vs. actual values for the entire data-set.

Test-$r^2$: This is correlation of predicted vs. actual during cross-validation by pooling omitted test cases from all the trials.

$$r^2 = 1 - \frac{\sum_{i=1}^{n}(y_i - y_{i(pred)})^2}{\sum_{i=1}^{n}(y_i - y_{mean})^2} \text{------------- (24)}$$

$$\text{RMSE} = \sqrt{\frac{\sum_{i=1}^{n}(y_i - y_{i(pred)})^2}{n}} \text{----------------- (25)}$$

### III. RESULTS AND DISCUSSION

The proposed fuzzy modeling approach is applied to a common benchmark problem: The antituberculosis prediction as a case study  In this typical nonlinear regression problem several attributes (input descriptors) are used to predict antituberculosis activity based on several given characteristics, such as molecular weight(MW), physicochemical properties such as logP, polar surface area, number of hydrogen bond donors and so on. The dataset is preprocessed before modeling.

The approximation power of the identified models is then compared with fuzzy models with the same number of rules obtained by the MATLAB fuzzy toolbox (ANFIS, neuro-fuzzy model) and the Fuzzy Model IDentification (FMID) toolbox that is based on Gustafson- Kessel clustering .

A TS-fuzzy model, that utilizes information profile data about the Oxazolines and Oxazoles derivatives, has been identified by the proposed clustering algorithm. The descriptors are : $u_1$: logp, $u_2$: polar surface area, $u_3$: number of hydrogen donors, $u_4$: molecular weight and so on.

The rules for the fuzzy model were developed based on the clustering of the data space and can be interpreted to reveal the nature of the relationships between activiy tthe descriptors and the    and even some information about the system. When $X_k = \phi_k = u_k$, the proposed clustering method gave RMSE values of  0.1540  and  0.4276 as well as $r^2$ values of 0.9638 and 0.8259  training and test set. Fig 1 is the scatter plot of the proposed model, which shows a correlation between observed value and antituberculosis activity  prediction in the training and test set.

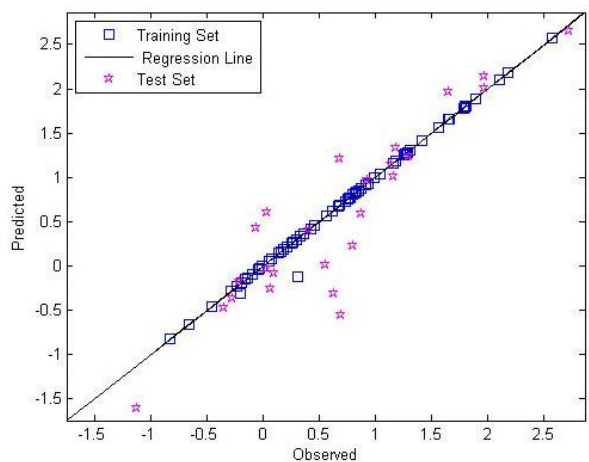

Figure 1. Correlation between observed and predicted values for training set and test set for proposed model(OLS and Fuzzy Identfication System)

The above presented model reduction techniques, OLS and FIS, are now applied to remove redundancy and simplify the model. First, based on the obtained fuzzy clusters, the OLS method is used for ordering of the input descriptors. The fuzzy model with two rules is applied. It turned out that for both clusters (local models) the "logp" and the "Polar surface area" become the most relevant input descriptors.

The FMID toolbox gives moderate results; RMSE values of 4.3558e-015 and 0.5576 as well as $r^2$ values of 1 and 0.7461 for train and test set. Fig 2 is the scatter plot of the FMID toolbox, which shows a correlation between observed value and antituberculosis activity prediction in the training and test set.

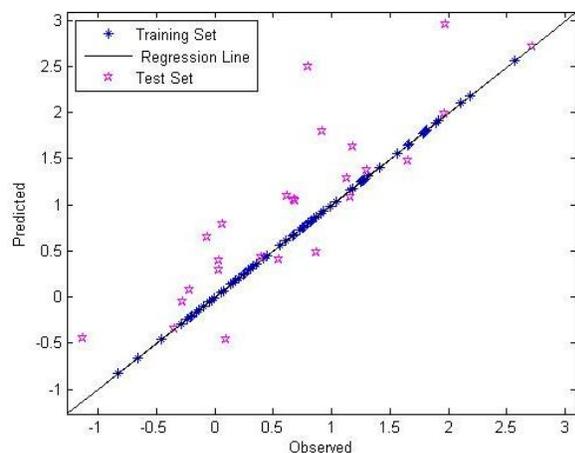

Figure 2. Correlation between observed and predicted values for training set and test set for FMID Toolbox

Bad results where obtained by the ANFIS algorithm when a neuro-fuzzy model with the same complexity was identified. RMSE values of 2.9305735220539218E-8 and 0.4608 as well as $r^2$ values of 1 and 0.7073 for train and test set. Fig 3 is the scatter plot of the ANFIS algorithm, which shows a correlation between observed value and antituberculosis activity prediction in the training and test set.

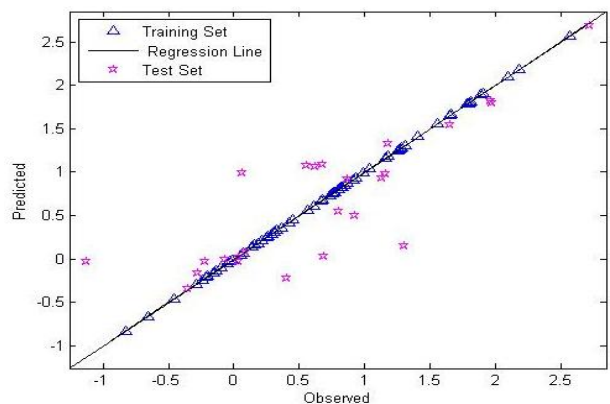

Figure 3. Correlation between observed and predicted values for training set and test set for ANFIS algorithm

Table 1 shows the performance of the proposed model and FMID tool box as well as ANFIS algorithm.

TABLE 1. PERFORMANCE OF THE IIDENTIFIED TS MODELS WITH INPUT DESCRIPTORS. HYBRID: CLUSTERING + OLS+ FIS, , ANFIS: NEURO FUZZY MODEL, FMID: FUZZY MODEL IDENTIFICATION TOOLBOX

| Method | Train RMSE | Test RMSE | Train $r^2$ | Test $r^2$ |
|---|---|---|---|---|
| Hybrid | 0.1540 | 0.4276 | 0.9638 | 0.8259 |
| FMID | 4.3558e-015 | 0.5576 | 1.0000 | 0.7461 |
| ANFIS | 2.9305735220539218E-8 | 0.4608 | 1.0000 | 0.7073 |

IV. CONCLUSION

We discussed the structure identification of Takagi-Sugeno fuzzy models and focused on methods to obtain compact but still accurate models. A new clustering methods is proposed to identify the local models and the antecedent part of the TS-fuzzy model. In addition two techniques for the selection of the antecedent and consequent attributes have been explained. Those techniques allow for the derivation of very compact models through subsequent ordering and removal of redundant and non-informative antecedent and consequent attributes. The approach is demonstrated by means of the prediction of the antituberculosis activity of Oxazolines and Oxazoles derivatives benchmark problem. The results show that the proposed approach is able to identify compact and accurate models in a straight forward way. This method is attractive in comparison with other iterative schemes, like FMID toolbox and ANFIS algorithm.

ACKNOWLEDGMENT

We gratefully thank to the Department of Computer Science Mangalore University , Mangalore India for technical support of this research.

REFRENCES